\documentclass[runningheads]{llncs}
\usepackage[T1]{fontenc}
\usepackage{graphicx}
\DeclareUnicodeCharacter{202F}{\,}

\usepackage{physics}

\begin{document}
\title{Air pollution models in epidemiologic studies \\ with geostatistics and machine learning}
%
%
\author{Manuel Ribeiro\inst{1}\orcidID{0000-0002-7890-7708}} 
\authorrunning{Manuel Ribeiro}
\titlerunning{Air pollution models with  geostatistics \& machine learning}
%
\institute{CERENA, Instituto Superior Técnico, Universidade de Lisboa\\
Av. Rovisco Pais, 1049-001 Lisboa, Portugal\\
\email{manuel.ribeiro@tecnico.ulisboa.pt}}
\maketitle              
\begin{abstract}
Development of air pollution models for large regions is a priority for population-based epidemiologic studies. The rapid development of big data information systems and machine learning algorithms have opened new grounds for refinements of current model frameworks. This commentary overviews recent contributions and outlines extensions from geostatistics and machine learning perspectives. For the coming years, expected advances will expand the use of learning algorithms to model spatial trends and incorporate spatial covariance models in the learning processes. These extensions will refine existing modelling frameworks contributing to improve accuracy of air pollution models for exposure assessment.

\keywords{geostatistics  \and machine learning\and air pollution \and exposure}
\end{abstract}
\section{Overview}\label{over}
Cancers, respiratory and cardiovascular diseases have been the most common reasons for premature death attributable to air pollution around the globe.
Still, apart from few exceptions, there is not a definite knowledge about the impacts of air pollution on populations. Therefore the development of accurate models predicting air pollution is critical to better understand the health impacts of exposure to air pollution.

Air pollution models require data collection, which often comes from monitoring by local or national institutions \cite{e16}, using a set of ground-based stations geographically dispersed over locations selected for regulatory purposes. Although these provide air pollutants information with high accuracy and high temporal resolution, they are sparse in space and have limited area coverage. In such cases, monitoring data can be combined with statistical models predicting pollution gradients between ground-based monitoring stations and in areas where no stations exist at all \cite{scl20}. To this end, remotely sensed data \cite{dssswsprsk19}, chemical transport models \cite{jtbidmscgkdctb17} or Copernicus Atmospheric Monitoring Service (CAMS) products \cite{sblb21} have been used. Similarly, meteorological data \cite{ybczgf17} and other auxiliary information such as land-use data or distance to roads are useful for modelling, predict and describe smaller-scale variations \cite{cglylmwnksyckj20,s18} and are a solution to obtain high-resolution air pollution maps \cite{scsgfprfv21,xbwsvbm19}.

More traditional modelling approaches are supported on linear combinations of potential predictors while considering their geographic locations. More specifically, land use regressions \cite{khpbv19,mg18}, employ linear regression combining geographic data collected on air pollutants and on potential predictors, and are a relatively cheap and practical approach to predict air pollution exposure in urban areas; spatial regressions \cite{zlzsfx18,zffz19}, allow linear regression parameters to vary smoothly as a function of spatial neighborhoods and increase the potential to capture non-stationary relations between predictors and air pollutants; geostatistical models \cite{rplbp16,xbwsvbm19}, take into account for spatial trends and spatial autocorrelation, capturing the intensity and direction of the spatial processes underlying air pollution concentrations, which is especially relevant in large regions, where variations in their intensity and direction are likely to occur.

In the last decade, approaches integrating machine learning techniques have been applied to handle non-linear interactions between predictors \cite{bjmo}, and hybrid models have been developed to account for spatial dependence of air pollutants \cite{smwb20}. Mostly focused on supervised learning, some popular algortihms used are decision trees \cite{gwlzp21}, random forest \cite{mbwzyhlsl21,zldzzgd18}, artificial neural networks \cite{abats20}.  Compared to the abovementioned linear models, the superiority of these algorithms relies in their capability to deal better with such complex non-linearities \cite{llzlj22}.

In addition to air pollution mapping, assessment of spatial uncertainty of predictions (hereinafter referred to as "uncertainty") is also required, since predictions have uncertainty that typically varies spatially and temporally. In fact, quantifying uncertainty (e.g. prediction intervals) provides important information about the prediction error in the air pollutant values used in exposure assessment, and addresses spatial misalignment of pollutant and health data \cite{gpzsc09}. Not taking uncertainty into account may produce misleading conclusions about the potential impacts on population´s health and weaken the scientific validity of its findings. In Bayesian framework, uncertainty can be assessed using a hierarchical formulation \cite{bv21}, or with geostatistical algorithms using stochastic modelling \cite{rp18}. Yet, quantification of spatial uncertainty is still in its infancy.

After this first section with an overview, the commentary is divided into 2 sections (second, third): section \ref{mod} presents some basics on geostatistical and machine learning and section \ref{geolearn} present some possible ways ahead in the perspective of combining geostatistical and machine learning methods. 

\section{Models} \label{mod}
While geostatistics is widely used in environmental applications of spatial data modelling, it shows at the same time, difficulties to model non-linear and complex dependency structures. Similarly, machine learning algorithms are powerful tools to solve complex real-world problems. Yet, they usually do not consider sample locations and spatial autocorrelation in the learning process. Combining the strengths of both approaches extends the existing methods to model air pollution, which is challenging due to the complex non-linear physical and chemical underlying processes and interactions affecting air pollution concentrations at different spatiotemporal scales.

\subsection{Geostatistical models}

Geostatistics includes a set of statistical techniques suited to model spatially correlated data, providing optimal unbiased predictions with minimum mean squared prediction error. Geostatistics are based on the assumption that sample data are a single realization of a spatial random process, $Y(\vb{s})$. In (\ref{eqn:geoY}), $S$ represents a finite spatial continuous domain and $\vb{s}$ is a spatial index.

\begin{equation}
\label{eqn:geoY}
\{Y(\vb{s}) : \vb{s} \in S\}
\end{equation}

To model spatially correlated data variogram functions are used, and Kriging (univariate case) or Cokriging (multivariate case) estimators are used for optimal linear prediction. There are different types of Kriging techniques, and the choice depends on data available and the aims of analysis. For a comprehensive review about Kriging methods please refer to \cite{g97}.

When the the objective of analysis is to assess spatial uncertainty, Kriging techniques are not adequate. Specifically, Kriging techniques aim at minimizing the prediction error, and this involves smoothing data variability. To assess spatial uncertainty, geostatistical algorithms using stochastic simulations should be preferred, as they reproduce the fluctuations observed in the sample data, instead of producing optimal prediction. These algorithms generate simulated maps with similar statistical properties to those of the observed data (e.g., histogram, spatial covariance), and have been applied to quantify air quality uncertainty to assess health impacts of exposure \cite{rp18,rplbp16,ygykd09}.

\subsection{Machine learning models}

In the past decade, many different learning algorithms have been applied to model air pollution for epidemiology studies \cite{bjmo}. Currently, these models use modelling frameworks with high dimensional input spaces including features such as air pollution (e.g., remotely sensed data, CTM, ground-based stations), meteorological-based models (e.g., temperature, wind, pressure, humidity), other relevant biophysical and socio-economic information (e.g., land-use land cover, seasonality, elevation, distance to roads, population density). 

Mostly focused on supervised learning regression algorithms, the learning processes use the input features, $\vb{x}$, for training and choose a function $f(\vb{x})$ with parameters $\vb{w}$, that better fits the output values $y$. To decide which function $f(\vb{x,w})$ provides the best approximation to $y$, a measure of loss is computed. A popular choice for regression problems is to tune $\vb{w}$ from data with the following loss function, $L$:

\begin{equation}
\label{eqn:L}
L[y, f(\vb{x,w})] = [y- f(\vb{x,w})]^2
\end{equation}

Random forest are one of the most used algorithms to predict air pollution \cite{rz18}, as they provide an excellent trade-off between interpretability and performance, when compared with other learning methods (e.g., neural networks, support vector machine). Simpler learning methods such as linear regression approaches have a long tradition in science and are also widely used \cite{llzlj22}. Following the principle of Occam's razor, when no major interactions and approximately linear associations exist, these models perform well and should be preferred instead of more complex algorithms. Neural networks are more complex than the abovementioned algorithms and are more efficient to solve non-linear problems. Several algorithmic variants are well established (e.g., artificial, convolutional neural networks)  in air pollution modelling \cite{sjkmk21}. In complex non-linear and high dimensional feature spaces, this algorithm can achieve high model performances \cite{cch19}.

The trade-off between model performance and the ability to generalize is a relevant issue to be taken into consideration, as the models should avoid overfitting the data (leading to low performance on new data). Moreover, the better performance of some complicated models may be hard to interpret. For example, in neural networks or in support vector machine algorithms, transfer functions and kernels used to fit the data, are rather artificial models and hard to interpret. A clearer understanding of the internal mechanisms leading to some output will contribute to improve their use in environmental applications. 

\section{Geostatistical learning models} \label{geolearn}

A modelling framework combining geostatistical and machine learning methods and explored in recent years, relies on the grounds of supervised machine learning with Regression Kriging \cite{hhs04}, also known as Kriging with External Drift (KED). Typically, KED decomposes a spatial random process $Y(\vb{s})$ into a function representing the deterministic part of the variation (e.g. linear model), $m(\vb{s})$, and a stochastic residual, $r(\vb{s})$, describing the spatially dependent part of variability:

\begin{equation}
\label{eqn:ked}
Y(\vb{s})= m(\vb{s}) + r(\vb{s})
\end{equation}

Real-world environments are highly non-linear and exhibit spatially dependent underlying processes. Therefore, simplistic linear model assumptions like linearity and independency are a limitation of traditional KED technique. The performance of the deterministic part, $m(s)$, can be improved with machine learning algorithms, as they do not require normally distributed data and are able to handle interactions and non-linear relationships between input features. In regression settings, random forests \cite{mbwzyhlsl21} or neural networks \cite{sjkmk21} are some of the algorithmic models that could be applied to model the deterministic part of RK.

\subsubsection{Uncertainty}

The spatial covariance model of the residual part, $r(s)$ can be inferred from data and incorporated in a geostatistical simulation algorithm to predict (map) spatially continuous surface of air pollution residuals. In fact, these simulation algorithms do characterize the spatial parameters of interest providing the means to generate realizations that reproduce spatial anisotropic correlation structure, and the empirical histogram of the stochastic residuals, $r(s)$.
Then, simulated results are added to the learned/tuned model outputs to create the final simulated air pollutant maps. Two key maps may be drawn from the set of geostatistical simulations: a pointwise median map of the variable of interest and the spatial uncertainty attached, which can be quantified by the pointwise interquartile range (IQR).

Parallelization and multithreading processes can be used as optimization techniques to run geostatistical simulation algorithms. In fact, these algorithms provide a measure of spatial uncertainty but at a computational high cost \cite{gkg11}. Several geostatistical simulation algorithms are available to assess spatial uncertainty of predictions (e.g., Sequential Gaussian Simulation, Direct Sequential Simulation, or Turning Bands). The performance of different algorithms can be measured using some loss-function, to select the best model. Readers may refer to Gómez-Hernandez and Srivastava \cite{gs21} for a comprehensive review of simulation algorithms. 

\subsubsection{Further extensions}

The ability to model and predict accurately air pollution with spatial data can be further refined by extending research in the perspective of machine learning tunning parameters. It is well known that air pollution exhibits spatial trends and spatial autocorrelation \cite{pmrrjpp20}. Relying on learning approaches based on random samples of spatial data to tune model parameters, fails to assess a model's performance in terms of spatial mapping and only validate its ability to reproduce sampling data \cite{mrwn19}. Therefore, instead of leaving spatial interactions to be learned from data, spatial parameters or functions (e.g., geostatistical semi-variogram) could be used as tuning parameter within the algorithmic model to control the optimal space and time ranges to sample \cite{hzdz21}. The idea can easily be extended to models aiming at predicting in the multipollutant case \cite{ss22}, by extending the spatial tunning parameters to cross-variograms. 

The influential work of Kanevski \cite{k09} offers a solid perspective on the grounds of machine learning developments for analysis and modelling of spatial environmental data, that can be easily transferable to air pollution modelling. In future work, researchers could extend the model framework, considering refinements in the assessment of spatial uncertainty, and optimization of parameters tunning by incorporating spatial properties of data.

\subsubsection{Aknowledgements} Manuel Ribeiro acknowledges Fundação para a Ciência e Tecnologia for the research contract IF2018/CP1384/IST-ID/175/2018 and CERENA pluriannual funding FCT-UIDB/04028/2020.
%
%
\bibliographystyle{splncs04}
\bibliography{mribeiro_libtex}

\begin{thebibliography}{10}
\providecommand{\url}[1]{\texttt{#1}}
\providecommand{\urlprefix}{URL }
\providecommand{\doi}[1]{https://doi.org/#1}

\bibitem{abats20}
Araujo, L.N., Belotti, J.T., Alves, T.A., Tadano, Y.d.S., Siqueira, H.:
  Ensemble method based on artificial neural networks to estimate air pollution
  health risks. Environmental Modelling and Software  \textbf{123}(10456), ~7
  (2020), \url{https://doi.org/10.1016/J.ENVSOFT.2019.104567}

\bibitem{bjmo}
Bellinger, C., Jabbar, M., M.~S., Z., O., O.V.: A systematic review of data
  mining and machine learning for air pollution epidemiology. BMC Public Health
   (2017), \url{https://doi.org/10.1186/s12889-017-4914-3}

\bibitem{bv21}
Beloconi, A., Vounatsou, P.: Substantial reduction in particulate matter air
  pollution across europe during 2006-2019: A spatiotemporal modeling analysis.
  Environmental Science and Technology  \textbf{55},  15505--15518 (2021),
  \url{https://doi.org/10.1021/acs.est.1c03748}

\bibitem{cch19}
Cabaneros, S.M., Calautit, J.K., Hughes, B.R.: A review of artificial neural
  network models for ambient air pollution prediction. Environmental Modelling
  and Software  (2019), \url{https://doi.org/10.1016/j.envsoft.2019.06.014}

\bibitem{cglylmwnksyckj20}
Cai, J., Ge, Y., Li, H., Yang, C., Liu, C., Meng, X., Wang, W., Niu, C., Kan,
  L., Schikowski, T., Yan, B., Chillrud, S.N., Kan, H., Jin, L.: Application of
  land use regression to assess exposure and identify potential sources in
  pm2.5, bc, no2 concentrations. Atmospheric Environment  \textbf{223} (2020),
  \url{https://doi.org/10.1016/j.atmosenv.2020.117267}

\bibitem{e16}
for Europe, W.R.O.: Health risk assessment of air pollution – general
  principles. Copenhagen (2016)

\bibitem{gwlzp21}
Gao, Y., Wang, Z., Li, C.y., Zheng, T., Peng, Z.R.: Assessing neighborhood
  variations in ozone and pm2.5 concentrations using decision tree method.
  Building and Environment  \textbf{188} (2021),
  \url{https://doi.org/10.1016/j.buildenv.2020.107479}

\bibitem{g97}
Goovaerts, P.: Geostatistics for Natural Resources Evaluation. Oxford
  University Press, New York (1997)

\bibitem{gpzsc09}
Gryparis, A., Paciorek, C.J., Zeka, A., Schwartz, J., Coull, B.A.: Measurement
  error caused by spatial misalignment in environmental epidemiology.
  Biostatistics  \textbf{10},  258–274 (2009),
  \url{https://doi.org/10.1093/biostatistics/kxn033}

\bibitem{gkg11}
Guan, Q., Kyriakidis, P.C., Goodchild, M.F.: A parallel computing approach to
  fast geostatistical areal interpolation. International Journal of
  Geographical Information Science  \textbf{25},  1241–1267 (2011),
  \url{https://doi.org/10.1080/13658816.2011.563744}

\bibitem{gs21}
Gómez-Hernández, J.J., Srivastava, R.M.: One step at a time: The origins of
  sequential simulation and beyond. Mathematical Geosciences  \textbf{53},
  193–209 (2021), \url{https://doi.org/10.1007/s11004-021-09926-0}

\bibitem{hhs04}
Hengl, T., Heuvelink, G.B.M., Stein, A.: A generic framework for spatial
  prediction of soil variables based on regression-kriging. Geoderma
  \textbf{120},  75–93 (2004),
  \url{https://doi.org/10.1016/j.geoderma.2003.08.018}

\bibitem{hzdz21}
Hoffimann, J., Zortea, M., de~Carvalho, B., Zadrozny, B.: Geostatistical
  learning: Challenges and opportunities. arXiv (arXiv:2102.08791) (2021)

\bibitem{dssswsprsk19}
de~Hoogh, K., Saucy, A., Shtein, A., Schwartz, J., West, E.A., Strassmann, A.,
  Puhan, M., Roösli, M., Stafoggia, M., Kloog, I.: Predicting fine-scale daily
  no2 for 2005-2016 incorporating omi satellite data across switzerland.
  Environmental Science and Technology  (2019),
  \url{https://doi.org/10.1021/acs.est.9b03107}

\bibitem{jtbidmscgkdctb17}
Jerrett, M., Turner, M.C., Beckerman, B.S., Iii, C.A.P., Donkelaar, A.v.,
  Martin, R.v., Serre, M., Crouse, D.L., Gapstur, S.M.S., Krewski, D., Diver,
  W.R., Coogan, P.F.P., Thurston, G.D., Burnett, R.T.: Comparing the health
  effects of ambient particulate matter estimated using ground-based versus
  remote sensing exposure estimates. Environmental Health Perspectives
  \textbf{125},  552–559 (2017)

\bibitem{k09}
Kanevsky, M.: Machine Learning for Spatial Environmental Data: Theory,
  Applications, and Software. EPFL Press (2009)

\bibitem{khpbv19}
Kerckhoffs, J., Hoek, G., Portengen, L., Brunekreef, B., Vermeulen, R.C.H.:
  Performance of prediction algorithms for modeling outdoor air pollution
  spatial surfaces. Environmental Science and Technology  \textbf{53},
  1413–1421 (2019), \url{https://doi.org/10.1021/acs.est.8b06038}

\bibitem{llzlj22}
Liu, X., Lu, D., Zhang, A., Liu, Q., Jiang, G.: Data-driven machine learning in
  environmental pollution: Gains and problems. Environmental Science and
  Technology  (2022), \url{https://doi.org/10.1021/acs.est.1c06157}

\bibitem{mbwzyhlsl21}
Ma, R., Ban, J., Wang, Q., Zhang, Y., Yang, Y., He, M.Z., Li, S., Shi, W., Li,
  T.: Random forest model based fine scale spatiotemporal o3 trends in the
  beijing-tianjin-hebei region in china, 2010 to 2017. Environmental Pollution
  \textbf{276} (2021), \url{https://doi.org/10.1016/j.envpol.2021.116635}

\bibitem{mrwn19}
Meyer, H., Reudenbach, C., Wöllauer, S., Nauss, T.: Importance of spatial
  predictor variable selection in machine learning applications – moving from
  data reproduction to spatial prediction. Ecological Modelling  \textbf{411}
  (2019), \url{https://doi.org/10.1016/j.ecolmodel.2019.108815}

\bibitem{mg18}
Morley, D.W., Gulliver, J.: A land use regression variable generation,
  modelling and prediction tool for air pollution exposure assessment.
  Environmental Modelling and Software  \textbf{105},  17–23 (2018),
  \url{https://doi.org/10.1016/j.envsoft.2018.03.030}

\bibitem{pmrrjpp20}
Pak, U., Ma, J., Ryu, U., Ryom, K., Juhyok, U., Pak, K., Pak, C.: Deep
  learning-based pm2.5 prediction considering the spatiotemporal correlations:
  A case study of beijing, china. Science of The Total Environment
  \textbf{699}(13356), ~1 (2020),
  \url{https://doi.org/10.1016/J.SCITOTENV.2019.07.367}

\bibitem{rp18}
Ribeiro, M.C., Pereira, M.J.: Modelling local uncertainty in relations between
  birth weight and air quality within an urban area: combining geographically
  weighted regression with geostatistical simulation. Environmental Science and
  Pollution Research  \textbf{25–54},  25942 (2018),
  \url{https://doi.org/10.1007/s11356-018-2614-x}

\bibitem{rplbp16}
Ribeiro, M.C., Pinho, P., Llop, E., Branquinho, C., Pereira, M.J.:
  Geostatistical uncertainty of assessing air quality using
  high-spatial-resolution lichen data: A health study in the urban area of
  sines, portugal. Science of the Total Environment  \textbf{562},  740–750
  (2016), \url{https://doi.org/10.1016/j.scitotenv.2016.04.081}

\bibitem{rz18}
Rybarczyk, Y., Zalakeviciute, R.: Machine learning approaches for outdoor air
  quality modelling: A systematic review. Applied Sciences (Switzerland)
  \textbf{8} (2018), \url{https://doi.org/10.3390/app8122570}

\bibitem{sblb21}
Schratz, P., Becker, M., Lang, M., Brenning, A.: Mlr3spatiotempcv:
  Spatiotemporal resampling methods for machine learning in r. arXiv
  \textbf{2110.12674} (2021)

\bibitem{sjkmk21}
Shams, S.R., Jahani, A., Kalantary, S., Moeinaddini, M., Khorasani, N.: The
  evaluation on artificial neural networks (ann) and multiple linear
  regressions (mlr) models for predicting so2 concentration. Urban Climate
  \textbf{37}(10083), ~7 (2021),
  \url{https://doi.org/10.1016/J.UCLIM.2021.100837}

\bibitem{smwb20}
Shao, Y., Ma, Z., Wang, J., Bi, J.: Estimating daily ground-level pm2.5 in
  china with random-forest-based spatiotemporal kriging. Science of the Total
  Environment  \textbf{740} (2020),
  \url{https://doi.org/10.1016/j.scitotenv.2020.139761}

\bibitem{scsgfprfv21}
Silibello, C., Carlino, G., Stafoggia, M., Gariazzo, C., Finardi, S., Pepe, N.,
  Radice, P., Forastiere, F., Viegi, G.: Spatial-temporal prediction of ambient
  nitrogen dioxide and ozone levels over italy using a random forest model for
  population exposure assessment. Air Quality, Atmosphere \& Health
  \textbf{14},  817–829 (2021),
  \url{https://doi.org/10.1007/s11869-021-00981-4/Published}

\bibitem{s18}
Son, Y., Osornio-vargas, A., Neill, M., Hystad, P., Texcalac-sangrador, J.,
  Ohman-strickland, P., Meng, Q., Schwander, S.: Land use regression models to
  assess air pollution exposure in mexico city using finer spatial and temporal
  input parameters. Science of the Total Environment  \textbf{639},  40 –48
  (2018), \url{https://doi.org/10.1016/j.scitotenv.2018.05.144}

\bibitem{ss22}
Song, J., Stettler, M.E.J.: A novel multi-pollutant space-time learning network
  for air pollution inference. Science of The Total Environment
  \textbf{811}(15225), ~4 (2022),
  \url{https://doi.org/10.1016/J.SCITOTENV.2021.152254}

\bibitem{scl20}
Sorek-Hamer, M., Chatfield, R., Liu, Y.: Review: Strategies for using
  satellite-based products in modeling pm2.5 and short-term pollution episodes.
  Environment International  \textbf{144}(10605), ~7 (2020),
  \url{https://doi.org/10.1016/j.envint.2020.106057}

\bibitem{xbwsvbm19}
Xu, H., Bechle, M.J., Wang, M., Szpiro, A.A., Vedal, S., Bai, Y., Marshall,
  J.D.: National pm2.5 and no2 exposure models for china based on land use
  regression, satellite measurements, and universal kriging. Science of the
  Total Environment  \textbf{655},  423–433 (2019),
  \url{https://doi.org/10.1016/j.scitotenv.2018.11.125}

\bibitem{ygykd09}
Young, L.J., Gotway, C.A., Yang, J., Kearney, G., DuClos, C.: Linking health
  and environmental data in geographical analysis : It's so much more than
  centroids. Spatial and Spatio-temporal Epidemiology  \textbf{1},  73–84
  (2009), \url{https://doi.org/10.1016/j.sste.2009.07.008}

\bibitem{ybczgf17}
Yáñez, M.A., Baettig, R., Cornejo, J., Zamudio, F., Guajardo, J., Fica, R.:
  Urban airborne matter in central and southern chile: Effects of
  meteorological conditions on fine and coarse particulate matter. Atmospheric
  Environment  \textbf{161},  221–234 (2017),
  \url{https://doi.org/10.1016/j.atmosenv.2017.05.007}

\bibitem{zlzsfx18}
Zhai, L., Li, S., Zou, B., Sang, H., Fang, X., Xu, S.: An improved
  geographically weighted regression model for pm2.5 concentration estimation
  in large areas. Atmospheric Environment  \textbf{181},  145–154 (2018),
  \url{https://doi.org/10.1016/j.atmosenv.2018.03.017}

\bibitem{zldzzgd18}
Zhan, Y., Luo, Y., Deng, X., Zhang, K., Zhang, M., Grieneisen, M.L., Di, B.:
  Exposure in china using hybrid random forest and spatiotemporal kriging
  model. Satellite-Based Estimates of Daily NO  \textbf{2} (2018),
  \url{https://doi.org/10.1021/acs.est.7b05669}

\bibitem{zffz19}
Zou, B., Fang, X., Feng, H., Zhou, X.: Simplicity versus accuracy for
  estimation of the pm2.5 concentration: a comparison between lur and gwr
  methods across time scales. Journal of Spatial Science p. 1–19 (2019),
  \url{https://doi.org/10.1080/14498596.2019.1624203}

\end{thebibliography}
%

\end{document}